\documentclass[preprint,showkeys,secnumarabic,showpacs,amsmath,amssymb]{revtex4}
\usepackage{graphicx}
\begin{document}

\title{On the Dynamics of Bianchi IX cosmological models}

\author{Hossein Farajollahi}
\email{hosseinf@guilan.ac.ir} \affiliation{Department of Physics,
University of Guilan, Rasht, Iran}

\author{Arvin Ravanpak}
\email{aravanpak@guilan.ac.ir}

\affiliation{Department of Physics, University of Guilan, Rasht,
Iran}

\date{\today}

\begin{abstract}
A cosmological description of the universe is proposed in the
context of Hamiltonian formulation of a Bianchi IX cosmology
minimally coupled to a massless scalar field. The classical and
quantum results are studied with special attention to the case of closed
Friedmann-Robertson-Walker model.
\end{abstract}



\maketitle

\section{Introduction}

 Any theoretical scheme of gravity must address a variety of conceptual
issues including the problem of time and identification of dynamical
observables  \cite{simeone:2}, \cite{alvarez:3},
\cite{carlini:4}, \cite{robert:5}, \cite{gogili:6}, \cite{marc:7},
\cite{marolf:8}, \cite{kitazo:9}, \cite{rafael:10},
 \cite{farajollahi:12}. Studying cosmological
models instead of general relativity helps us to overcome the
problems related to the infinite number of degrees of freedom in the
theory and pay more attention to the issues arising from the time
reparametrization invariance of the theory; such as the
identification of a dynamical time and also construction of
observables for the theory \cite{Farajollahi:11}, \cite{wald:13},
\cite{wood:14}, \cite{palli:15}. In particular, Bianchi IX
cosmological model is an interesting candidate to test the possible
solutions of the above problems \cite{montani:1}, \cite{sav:16},
\cite{donald:17}.

The problem of time is the most known difficulty of the
Wheeler-DeWitt (WDW) quantum geometrodynamics which is a theoretical
basis for modern quantum cosmology \cite{miller:18}, \cite{josh:19}.
Time in quantum mechanics and general relativity are drastically
different from each other; in quantum mechanics time is a global and
absolute parameter, in general relativity a local and dynamical
variable \cite{Wald:20}. As a result, the wave function in quantum
formulation is time independent, i.e., the universe has a static
picture.

A common candidate for a dynamical time in classical and quantum
model is a massless scalar field coupled to gravity
\cite{montani:1}, \cite{palli:15}, \cite{ramos:21}. In this work we
apply this to Bianchi IX cosmological model and investigate its
classical and quantum implementation. Due to the presence of the
minimally coupled scalar field term in the formulation, we show that
the dynamics of the metric functions can be obtained using a time
variable, as a function of the scalar field.

The article is organized as follows: In section two the Hamiltonian
formulation of Bianchi IX model coupled to a massless scalar field
is studied, with particular attention to the closed FRW model. In
section three the canonical quantization of the model is presented
and the wave function of the universe is discussed. Again special
attention is given to the quantum FRW cosmological solutions with
closed curvature. Section four presents conclusion drawn from this
work.

\section{The Classical Model}

The characteristic feature of the Bianchi IX model is the existence
of the simply transitive isometry group. The infinitesimal
generators of this group are the three linearly independent
spacelike killing vectors, $\xi_i,$ which obey
\begin{equation}
[ \xi_i , \xi_j ] = {C_{ij}}^k \xi_k,
\end{equation}
where $i$, $j$ and $k$ runs from 1 to 3. The generators of these
transformations are
\begin{eqnarray}
\xi_1 &=& x^2\partial_3 - x^3\partial_2, \nonumber\\
\xi_2 &=&  x^3\partial_1 - x^1\partial_3, \\
\xi_3 &=&  x^1\partial_2 - x^2\partial_1. \nonumber
\end{eqnarray}
One may write the metric of Bianchi IX in terms of {\it forms},
\begin{equation}
ds^2 = N^2dt^2 - h_{ij}\sigma^i\sigma^j,
\end{equation}
where $N$ is the lapse function, $h_{ij}$ is the 3-metric and the
$\sigma_i$ are the left invariant {\it 1-forms} of $SU(2)$:
\begin{eqnarray}
\sigma^1 &=& cos\psi d\theta + sin\psi sin\theta d\phi, \nonumber\\
\sigma^2 &=& -sin\psi d\theta + cos\psi sin\theta d\phi, \\
\sigma^3 &=& d\psi + cos\theta d\phi, \nonumber
\end{eqnarray}
with $0\leq \psi \leq 4\pi$  ,  $0\leq \theta \leq \pi$  and $0\leq
\phi \leq 2\pi$.

Without lose of generality, we can assume that the metric $h_{ij}$
is diagonal. Using the Misner variables to parameterize the metric
\begin{equation}
h_{ij} = e^{2\alpha} (e^{2\beta})_{ij},
\end{equation}
where $\alpha$ and $\beta$ are functions of time only. The matrix
\begin{equation}
\beta_{ij} = diag[\beta_+ + \sqrt3 \beta_- , \beta_+ - \sqrt3
\beta_- , -2\beta_+],
\end{equation}
with the property $Tr\beta = 0$, ensures that the 3-volume of the
hypersurface depends only on the conformal factor $\alpha$. The
variables $\beta_+$ and $\beta_-$ describe the anisotropy of the
spacetime, and, in particular, if they equal to zero, the model
reduces to the ordinary closed FRW model.

The Hilbert action for the model, including the massless scalar
field minimally coupled to gravity, is given by
\begin{equation}
S_{H} = \int \sqrt{-g}(R -
\frac{1}{2}g^{{\mu}{\nu}}\phi_{,\mu}\phi_{,\nu}) d^4x,
\end{equation}
where $\mu$ and $\nu$ runs from 1 to 4 and $\sqrt{-g}=N\sqrt{h}$. If
one assume that the scalar field is spatially homogeneous, the
Lagrangian for the model in terms of the Misner variables becomes
\begin{equation}
L = \frac {-6e^{3\alpha}}{N}(\dot\alpha^2 - \dot\beta_+^2 -
\dot\beta_-^2) + \frac{Ne^{\alpha}}{2} V(\beta_+ , \beta_-) -
\frac{e^{3\alpha}}{2N}{\dot\phi}^2,
\end{equation}
where
\begin{equation}
V(\beta_+ , \beta_-)= e^{-8\beta_+} -
4e^{-2\beta_+}\cosh(2\sqrt3\beta_-) +
2e^{4\beta_+}(\cosh(4\sqrt3\beta_-)-1).
\end{equation}
For the Hamiltonian formulation of the theory, let's determine the
conjugate momenta to the dynamical variables $\alpha$, $\beta_+$,
$\beta_-$, $\phi$, and $N$:
\begin{eqnarray}
&&P_\alpha = \frac{\partial L}{\partial\dot\alpha} =
\frac{-12e^{3\alpha}\dot\alpha}{N},\\
&&P_+ = \frac{\partial L}{\partial\dot\beta_+} = \frac{12e^{3\alpha}\dot\beta_+}{N},\\
&&P_- = \frac{\partial L}{\partial\dot\beta_-} = \frac{12e^{3\alpha}\dot\beta_-}{N},\\
&&P_ \phi = \frac{\partial L}{\partial\dot\phi} = -\frac{e^{3\alpha}\dot\phi}{N},\\
&&P_N = \frac{\partial L}{\partial\dot N} = 0.
\end{eqnarray}
Obviously, the variable  $N$ is not a canonical variable as its
canonical conjugate momenta is zero. The canonical Hamiltonian can
then be written as
\begin{equation}
H_c = \frac{e^{-3\alpha}N}{24} (-P_\alpha^2 + P_+^2 + P_-^2) -
\frac{Ne^{\alpha}V(\beta_+ , \beta_-)}{2} -
\frac{Ne^{-3\alpha}{P_\phi}^2}{2},
\end{equation}
with the total Hamiltonian to be
\begin{equation}
H_t = H_c + \lambda P_N.
\end{equation}
Since $P_N\approx 0$ is a primary constraint, preservation of this
constraint over time yields a secondary constraint (called
Hamiltonian constraint),
\begin{equation}
{\cal{H}} = \frac{e^{-3\alpha}}{24} [P_\alpha^2 - P_+^2 - P_-^2 +
12e^{4\alpha}V(\beta_+ , \beta_-)] +
\frac{e^{-3\alpha}{P_\phi}^2}{2} \approx 0.
\end{equation}
By choosing $N_1 = Ne^{-3\alpha}$ and introducing a new variable $a
= e^\alpha$, ($a$ is scale factor) we can redefine the secondary
constraint as
\begin{equation}
{\cal{H}}_1 = \frac{1}{24} [a^2P_a^2 - P_+^2 - P_-^2 +
12a^4V(\beta_+ , \beta_-)]+ \frac{{P_\phi}^2}{2} \approx 0,
\end{equation}
where $P_a=\frac{\partial L}{\partial \dot{a}}$. Thus
\begin{equation}
H_t = \lambda P_N + N_1{\cal H}_1,
\end{equation}
in which according to Dirac procedure, both of these constraints are
first class.

Now, we define a new time variable and its conjugate momenta as
\cite{Farajollahi:11}:
\begin{equation}
T= \int_{\Sigma_t}\frac{\phi}{P_\phi}\sqrt hd^3x,
\end{equation}
\begin{equation}
\Pi_T = \frac{P_\phi^2}{2},
\end{equation}
where the integration is over the spatial hypersurface $\Sigma_t$.
The Hamiltonian constraint in terms of the new variables becomes
\begin{equation}
{\cal {H}}_1 = {\cal{H}}^* + \Pi_T \approx 0,
\end{equation}
where ${\cal{H}}^*$ is the Hamiltonian constraint without scalar
field.

The equation of motion for $T$,
\begin{equation}
\frac{dT}{dt} = \int_{\Sigma_t}N_1\sqrt hd^3x,
\end{equation}
shows that the time variable equals the four-volume enclosed between
the initial and final hypersurfaces, which is necessarily positive
and monotonically increasing and can play the role of a cosmological
time. Even though, the time variable is not a Dirac observable, it
can be used to play the role of a cosmological time. Besides, the
conjugate momentum to this variable is a Dirac observable.

Classically, We can analyze in which sense the dynamical time, $T$,
can be regarded as an appropriate time variable for the theory. Near
the cosmological singularity $(a \rightarrow 0)$ the potential term
$V (\beta_+ , \beta_-)$ can be neglected and so we have
\begin{equation}
{\cal{H}}_1 = \frac{1}{24} [a^2P_a^2 - P_+^2 - P_-^2] + \Pi_T
\approx 0.
\end{equation}
In such approximation we obtain the new equations of motion
\begin{eqnarray}
&&a' = \frac{1}{12}a^2P_a, \\
&&{P_a}' = -\frac{1}{12}a{P_a}^2, \\
&&{P_+}' = {P_-}' = 0, \\
&&{\beta_+}' = -\frac{1}{12}P_+, \\
&&{\beta_-}' = -\frac{1}{12}P_-,
\end{eqnarray}
which , $(...)'$, denotes derivative of $(...)$, with respect to
$T$. The solutions to the above system  for the scale factor $a$ and
its conjugate momenta are
\begin{eqnarray}
a &=& a_0e^{CT},\\
P_a &=& P_0e^{-CT},
\end{eqnarray}
where $a_0 = a|_{T=0}$, $P_0 = P|_{T=0}$ and $C$ is a constant. A
positive $C$ with positive $a_0$ provides an accelerating expanded
universe, with a positive constant Hubble, $H = C$. We have
recovered a monotonic dependence of our new variable $T$ with
respect to the isotropic variable of the Universe $a$ and therefore
$T$ shows to be a relational time variable for the gravitational
dynamics. In this case, the universe expands exponentially according
to the cosmological time variable, and naturally accelerating
without a beginning singularity.

In the presence of the potential term, i.e., far from the
singularity, the equations of motion are
\begin{eqnarray}
a' &=& \frac{1}{12}a^2P_a, \\
{P_a}' &=& -\frac{1}{12}a{P_a}^2 -2a^3[e^{-8\beta_+} -
4e^{-2\beta_+}\cosh(2\sqrt3\beta_-)\nonumber \\
&+& 2e^{4\beta_+}(\cosh(4\sqrt3\beta_-)-1)], \\
{\beta_+}' &=& -\frac{1}{12}P_+ ,\\
{\beta_-}' &=& -\frac{1}{12}P_- ,\\
{P_+}' &=& -4a^4[-e^{-8\beta_+} +
e^{-2\beta_+}\cosh(2\sqrt3\beta_-)\nonumber \\
&+& e^{4\beta_+}(\cosh(4\sqrt3\beta_-)-1)], \\
{P_-}' &=& -4\sqrt3a^4[-e^{-2\beta_+}\sinh(2\sqrt3\beta_-) +
e^{4\beta_+}\sinh(4\sqrt3\beta_-)].
\end{eqnarray}
Although, it is hard to solve these set of coupled nonlinear
differential equations analytically, a numerical solution of the
equations shows, in particular, the dynamic of the scale factor in
terms of the time variable $T$ (Fig.1). Again, We have recovered a
monotonic dependence of our new variable $T$ with respect to the
isotropic variable of the Universe $a$ and the universe expansion
begins with no singularity.
\begin{figure}
\centering
\includegraphics[scale=0.35]{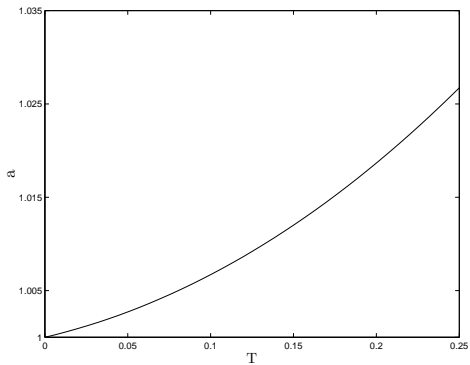}
\caption{dynamics of scale factor with respect to $T$}
\end{figure}

${\bf Isotropic \ Bianchi \ type\  IX}$:

In particular, we are more interested to find the behavior of the
isotropic variables $a$ and $p_a$ when $\beta_+=0$ and $\beta_-=0$.
We find the equations of motions as,
\begin{equation}
a'=\frac{a^2P_a}{12}\label{aprime},
\end{equation}
and
\begin{equation}
P'_a=-\frac{aP_a^2}{12}-6a^3\label{pprime}.
\end{equation}

The nontrivial solution of the above coupled non linear differential
equations~(\ref{aprime}) and (\ref{pprime}), for the scale factor
gives
\begin{equation}
a(T)=\frac{2\sqrt{c_1}e^{\sqrt{\frac{c_1}{2}}(T+c_2)}}{(4c_1+e^{2\sqrt{c_1}(T+c_2)})^{\frac{1}{2}}},
\end{equation}
where $c_1$ and $c_2$ are positive constants of integrations. This
solution shows that the universe has no singularity at all. Besides,
the universe shrinks to a big crunch as $T$ goes to infinity while
it reaches a maximum size during its history. Note that, in the isotropic limit, the Bianchi IX model
contains the closed FRW model and the behavior of the scale factor $a$ is as expected for this model.

\section{The Quantum Model}

In quantization of the theory, we obtain the Wheeler-DeWitt equation
\begin{equation}
\widehat{\cal H}_1\psi = 0,
\end{equation}
or
\begin{equation}
\widehat{{\cal {H}}^*}\psi = -\Pi_T\psi.
\end{equation}
So, we get a Schr\"{o}dinger like equation
\begin{equation}
\widehat{{\cal {H}}^*}\psi = i\hbar\frac{\partial}{\partial T}\psi,
\end{equation}
where the new variable plays the role of a dynamical time for the
theory. Explicitly, the Schr\"{o}dinger equation can be written as:
\begin{equation}
\frac{1}{24}[a^2\frac{\partial^2}{\partial
a^2}-\frac{\partial^2}{\partial
\beta_+^2}-\frac{\partial^2}{\partial \beta_-^2}]\psi =
i\hbar\frac{\partial}{\partial T}\psi,
\end{equation}
where we have ignored the potential term in this regime.

In fact, in quantum cosmology, the Universe is described by a single
wave function $\psi$ providing puzzling interpretations when
analyzing the differences between ordinary quantum mechanics and
quantum cosmology. If we consider the universe wave function as:
\begin{equation}
\psi(a, \beta_+, \beta_-; T) =
{\cal{A}}(a)B_+(\beta_+)B_-(\beta_-){\cal{T}}(T),
\end{equation}
then, by using separation of variables method we obtain
\begin{eqnarray}
  &&i\hbar\frac{1}{\cal{T}}\frac{d\cal{T}}{dT}=E,\\
  &&\frac{1}{B_+}\frac{d^2B_+}{d\beta_+^2}+m^2=0,\\
  &&\frac{1}{B_-}\frac{d^2B_-}{d\beta_-^2}+n^2=0,\\
  &&\frac{a^2}{\cal{A}}\frac{d^2\cal{A}}{da^2}+l^2=0,
\end{eqnarray}
where $l^2 = m^2 + n^2 + k^2$,  $k^2 = \frac{24\pi E}{\hbar^2}$ and
$E$, $k$, $m$, $n$ and $l$ are arbitrary constants. The solutions to
these equations are:
\begin{eqnarray}
  &&{\cal{T}}(T) = \exp(\frac{-iET}{\hbar}), \\
  &&B_+(\beta_+) =c_1\cos(m\beta_+) + c_2\sin(m\beta_+), \\
  &&B_-(\beta_-) = {c_1}'\cos(n\beta_-) + {c_2}'\sin(n\beta_-), \\
  &&{\cal{A}}(a) =\sqrt a( {c_1}''\cos(\frac{{\sqrt{(1-4l^2)}}\ln a}{2})
+ {c_2}''\sin(\frac{{\sqrt{(1-4l^2)}}\ln a}{2})),
\end{eqnarray}
where $c_1$, $c_2$, ${c_1}'$, ${c_2}'$, ${c_1}''$ and ${c_2}''$ are
normalisation coefficients.

For the Hamiltonian operator to be self adjoint the wave function
must satisfy one of the following boundary conditions
\cite{batista:22}, \cite{alvarenga:23}
\begin{eqnarray}
  &&\psi(a, \beta_+, \beta_-, T)|_{a=0}=0, \label{boundary1}\\
  &&\frac{\partial \psi(a, \beta_+, \beta_-, T)}{\partial a}|_{a=0} = 0. \label{boundary2}
\end{eqnarray}
The first boundary condition is satisfied while the second one leads
to infinity. Obviously, the wave function is not square integrable
and in order to obtain a possible physical solution we may construct
wave packets as~\cite{alvarenga:23}
\begin{equation}
\psi(a, \beta_+, \beta_-; T) = \int A(E)\psi_E(a, \beta_+,
\beta_-;T)dE.
\end{equation}
However, the wave packets are also problematic and a satisfactory
solution to the WDW equation is not easy to obtain.

${\bf Isotropic \ Bianchi \ type\  IX}$:

In the case of isotropic universe, the associated Wheeler-DeWitt
equation, i.e., the Schr\"{o}dinger-like equation is
\begin{equation}
-\partial_a^2 \Psi-36a^2\Psi -i\frac{24}{a^2}\partial_T\Psi=0.
\end{equation}
The solution is
\begin{equation}
\Psi_E(a, T)=e^{-iET}\sqrt{a}\left[c_3{J_+}_\nu(3a^2)+c_4{J_-}_\nu(3
a^2)\right],
\end{equation}
where $c_3$ and $c_4$ are constants of integrations, ${J_+}_\nu$ and
${J_-}_\nu$ are Bessel functions and $\nu=\sqrt{1-96E}/4$. Since
${J_-}_\nu$ grows exponentially to infinity as $a$ goes to zero, one
must set $c_4=0$, and consequently the first boundary
condition,~(\ref{boundary1}), is satisfied. In this isotropic case
the wave function is still not square integrable and even using wave
packets does not give us a satisfactory wave function for the
universe.

\section{Conclusion}

In this work,  we present the Hamiltonian formulation of the Bianchi
type IX cosmological model minimally coupled to a scalar field. We
show that the dynamics of the metric functions can be obtained using
a time variable, $T(t)$, as a function of the scalar field.

We then apply the results for the classical and quantum models. We
find that the classical model has solutions which avoid the usual
initial cosmological singularity. Particularly, in a positive
curvature spacetime, the solution also shows a big crunch in future.
In the quantum description of the model, however, a satisfactory wave function
is difficult to be physically interpreted. Even in the case of isotropic
Bianchi type IX model, it is not easy to find a normalisable solution to
the WDW equation.

\end{document}